\documentclass[aps,prl,reprint,nobibnotes,nofootinbib]{revtex4-1}
\usepackage[margin=1in]{geometry}
\usepackage{amsmath,amssymb,amsfonts,amsthm,mathtools}
\usepackage{bbm}
\usepackage{hyperref}
\usepackage{soul}
\hypersetup{colorlinks=true,linkcolor=blue,citecolor=blue,urlcolor=blue}
\usepackage{pgfplots}
\pgfplotsset{compat=1.18}
\usetikzlibrary{arrows.meta,positioning,calc}
\newtheorem{theorem}{Theorem}[section]
\newtheorem{lemma}[theorem]{Lemma}
\newtheorem{definition}[theorem]{Definition}

\newtheorem{proposition}[theorem]{Proposition}

\usepackage{tikz}
\usepackage[compat=1.1.0]{tikz-feynman}
\usetikzlibrary{arrows.meta,calc,decorations.pathmorphing,decorations.markings}

\newcommand{\R}{\mathbb{R}}
\newcommand{\C}{\mathbb{C}}

\newcommand{\supp}{\operatorname{supp}}
\newcommand{\norm}[1]{\left\|#1\right\|}
\newcommand{\braket}[2]{\left\langle #1 \middle| #2 \right\rangle}
\newcommand{\cH}{\mathcal{H}}

\newcommand{\id}{\mathbbm{1}}
\newcommand{\e}{\mathrm{e}}
\newcommand{\dd}{\mathrm{d}}

\begin{document}

\title{On the Meaning of Localization in Non-Local Quantum Field Theory}

\author{E.~J.~Thompson}
\thanks{Corresponding author.}
\email{Thom3471@mylaurier.ca}
\affiliation{Wilfrid Laurier University, Waterloo, Canada, N2L 3C5}

\date{\today}

\begin{abstract}
In this paper we explore and derive an uncertainty principle for an ultraviolet complete nonlocal quantum field theory where under our hypothesises of an induced equal time detector response kernel, we then prove that the observed localization width obeys an exact variance addition law. Then when we combine this with the ordinary Heisenberg inequality and we obtain a nonlocal uncertainty relation. The bound reduces to the usual local relation in the infrared or local limit when $E_M \to \infty$, while in the ultraviolet it implies a minimal localization length of order $L_M$. We go on to explain what this means for locality, microcausality, the interpretation of spacetime points, and the ultraviolet structure of quantum field theory. In this formulation we note and prove that spacetime will remain a Lorentz covariant continuum at the level of the manifold description but pointlike localization ceases to be a physically realizable observable notion below the nonlocality scale.
\end{abstract}

\maketitle

\section{Introduction}

The standard Heisenberg uncertainty principle is one of the most basic structural statements in quantum theory that says in its simplest form that a state cannot be simultaneously sharply localized in position and momentum, or to put it in another way we cannot simultaneously know the precise position and momentum of a particle~\cite{Heisenberg,Kennard,Robertson,Schrodinger}.  The more accurately one property is measured the less accurately the other can be known due to the inherent wave-like behavior of matter. The quantum hypothesis first entered modern physics with Planck’s black-body analysis~\cite{Planck1901}, and this wave-like behavior of matter was shown by de Broglie in 1924 following the proposed wave-particle duality of light proposed by Albert Einstein in 1905~\cite{Einstein1905,deBroglie1924}. Between 1914 and 1916 Robert Millikan worked trying to disprove Einstein’s theory of the photoelectric effect but instead his precise measurements ended up confirming it perfectly proving that light energy is indeed delivered in discrete quanta~\cite{Millikan1916}. Then the Compton Effect found in 1923 by Arthur who Compton shot X-rays at electrons and found that the X-rays bounced off like billiard balls, this showed that light doesn't just have energy but it also has momentum, a classic particle trait~\cite{Compton1923}. The language of the photon itself entered the literature shortly thereafter when it was defined by Lewis in 1926~\cite{Lewis1926Photon}. Then the work of de Broglie was experimentally validated shortly after by Clinton Davisson and Lester Germer, as well as George Paget Thomson in 1927, these researchers demonstrated that electrons can be diffracted by crystals and this showed us that they behave as waves~\cite{DavissonGermer1927,ThomsonReid1927}. Louis de Broglie was awarded the 1929 Nobel Prize in Physics for his discovery of the wave nature of electrons that were before thought to be simply particle like.

In relativistic quantum field theory (QFT) the meaning of localization becomes much more subtle~\cite{NewtonWigner,Wightman1962,Malament1996,Fleming1965,Miller1972,Hegerfeldt}. A full interacting QFT is not naturally organized around a fundamental position operator in the same way as nonrelativistic quantum mechanics but instead local QFT is organized by local or non-local observables, causal commutator relations, and measurement operations localized in spacetime regions~\cite{Haag1992,FewsterRejznerAQFT,BuchholzFredenhagenAQFT}. In this setting one must distinguish between at least three different notions, the first being the canonical one-particle uncertainty relation in a fixed representation, the second being localization defined by sharply supported projectors or idealized position observables~\cite{NewtonWigner,Wightman1962,Malament1996,Busch1999,Sorkin1993Impossible,FoldyWouthuysen,ThompsonOperational}, and third the localization defined by detector effects or POVMs induced by localized interactions~\cite{Busch1999,FewsterVerch,FewsterVerchMeasurement,PapageorgiouFraserMeasurement,FalconeContiLocalization}. A further reason for this caution we take is that in relativistic quantum theory with positive energy sharply localized states exhibit instantaneous spreading so naive localization concepts must be handled with care as shown in~\cite{Hegerfeldt}.

A measurement-theoretic viewpoint has a rich history and deep roots in the analysis of measurable field quantities dating back to the early 1900s over debates on how field theories should be though of~\cite{LandauPeierls1931,BohrRosenfeld,BohrRosenfeld2}

The distinction becomes essential in ultraviolet-complete non-local quantum field theory where physical observables are generated not by strictly point-supported fields but by regulated smearings involving an entire function of the d'Alembertian~\cite{PaleyWiener,Moffat1990FiniteNonlocalGauge,EvensMoffatKleppeWoodard1991,MoffatThompsonReg,ThompsonCovariance}. In this setting exact bounded-region projectors are no longer realizable, and localization acquires an intrinsic finite resolution scale.

Ultraviolet divergences in relativistic quantum field theory have motivated a long and broad search for frameworks beyond the strict local point field dynamics including S-matrix and non-pointlike alternatives~\cite{Dyson1949Radiation,Dyson1949SMatrix,Dyson1952Divergence,PaisUhlenbeck1950}. Early relativistically motivated cutoff proposals and the spacetime reformulation of quantum electrodynamics already made clear that ultraviolet modification and gauge structure must be handled together~\cite{Feynman1948Cutoff,Feynman1949QED}. There were many early attempts to formulate finite or extended nonlocal quantum field theories that go back to Yukawa’s nonlocal-field program, the work of Rivier and St\"uckelberg, the Nagoya
group discussion of vacuum polarization, Rayski’s nonlocal electrodynamics, Blokhintsev’s
nonlocal and nonlinear field theories, and later the mathematically developed nonlocal
program of Efimov and Alebastrov~\cite{Yukawa1950a,Yukawa1950b,RivierStueckelberg1948,UmezawaYukawaYamada1948,Rayski1951,Blokhintsev1958,Efimov1967,Efimov1970,AlebastrovEfimov1973,AlebastrovEfimov1974,Moffat1960RegVEV}.

The purpose of this paper is to derive from the structure of entire function regulated non-local quantum field theory itself, the corresponding modification of the uncertainty principle. The derivation we present is not an ad hoc postulate of a deformed canonical commutator but rather it begins with the actual non-local observable algebra and the associated measurement-theoretic notion of localization. The result is a non-local uncertainty principle. In particular the measured localization width is broadened by the non-local response kernel induced by the regulator, and this leads to an exact variance-addition law and hence a modified uncertainty relation. The resulting bound reduces to the usual Heisenberg relation in the infrared and yields a minimal non-local localization length in the ultraviolet.

The central conclusion is therefore not that spacetime coordinates necessarily become noncommutative, nor that spacetime must become discrete, but that pointwise localization ceases to be a physically implementable notion below the non-locality length $L_M \sim E_M^{-1}$~\footnote{As a tribute to the man who formulated this version of QFT that has brought forth so many interesting results and who taught me almost everything I know as a scientist we call this the Moffat length and the associated Moffat energy, and more generally the Moffat parameters if one were to extend it to time in terms of $t^2_M=\hbar G/c^5$, we note this as one might expect as when we have nonlocal length, mass, and energy we ought to have some notion of nonlocal time in the same way we do this with the Planck units~\cite{Planck1900IrreversibleRadiation,Planck1897IrreversibleI,Planck1897IrreversibleII,Planck1897IrreversibleIII,Planck1898IrreversibleIV,Planck1899IrreversibleV,Stoney1881PhysicalUnits,Tomilin1999NaturalUnits,Barrow1983NaturalUnits,Wilczek2005AbsoluteUnits,Meschini2007PlanckScale}.}. In this we see that exact microcausality and point-supported observables become emergent infrared ideals rather than fundamental objects, while the fundamental ultraviolet theory is nonlocal.

Early proposals for a UV complete theory include Wataghin's finite-radius form factors, Yukawa's theory of non-local fields, the causal and non-local interaction program of Rivier and Stueckelberg, Rayski's non-local electrodynamics, Blokhintsev's non-local field-theoretic constructions, and the systematic non-local quantum field theory program of Efimov and Alebastrov~\cite{Wataghin1934,Yukawa1950,Yukawa1950b,RivierStueckelberg1948,StueckelbergRivier1950,Rayski1951,Blokhintsev1947,Efimov1967,AlebastrovEfimov1973,Efimov1974}. The exponential and entire function form factors used in modern non-local field theory therefore have a long history and should not be regarded as new ingredients of this paper~\cite{PaisUhlenbeck1950,Efimov1967,Efimov1974}.

Later developments include non-local scalar, gauge, and gravitational field theories studied by Krasnikov, Kuz'min, Tomboulis, Moffat, Kleppe, Woodard, and others~\cite{Krasnikov1987,Kuzmin1989,Tomboulis1997,Moffat1990,EvensMoffatKleppeWoodard1991,KleppeWoodard1992}. More recent work on weakly non-local or infinite-derivative field theory and gravity includes contributions by Biswas, Koshelev, Briscese, Buoninfante, Calcagni, Mazumdar, Modesto, Rachwa{\l}, Shapiro, and collaborators~\cite{BiswasMazumdarSiegel2006,Koshelev2007,BiswasGerwickKoivistoMazumdar2012,Modesto2012,BrisceseModestoTsujikawa2014,ModestoRachwal2014,CalcagniModesto2014,CalcagniModesto2015,ModestoRachwal2015,BuoninfanteLambiaseMazumdar2019,BuoninfanteGhoshalLambiaseMazumdar2019,BrisceseModesto2019}. Recent handbook reviews now provide very useful entry points to this literature~\cite{KrasnikovHandbook2023,BasBeneitoCalcagniRachwal2023,BuoninfanteGiacchiniNettoHandbook2023}.

To state what is and is not new in thus paper we note that non-local field theories, exponential or entire function form factors, non-local gauge theories, non-local gravity, Fock-space constructions, expectation values, smeared observables, and convolution kernels have all appeared extensively in the literature~\cite{Yukawa1950,PaisUhlenbeck1950,Efimov1967,AlebastrovEfimov1973,Krasnikov1987,Kuzmin1989,Tomboulis1997,Moffat1990,EvensMoffatKleppeWoodard1991,KleppeWoodard1992,BiswasMazumdarSiegel2006,Modesto2012,ModestoRachwal2015}. The present paper does not claim novelty in any of these ingredients.

The new point we wish to make is the derivation of an uncertainty relation from the operational localization structure of the regulated theory, in this paper we assume that the physical localization observable is represented by a detector response kernel\footnote{The detector language is used here only as an way of displaying the finite resolution structure of the theory, but the kernel \(\kappa_F\) should not be interpreted as an arbitrary apparatus dependent imperfection. It is induced by the same entire-function regulator \(F(\Box/E_M^2)\) that defines the non-local quantum field theory. Equivalently, we can formulate the argument directly at the level of the regulated observable algebra where any local density \(\mathcal N(t,\mathbf x)\), such as a charge density, energy density, or one-particle number density in an appropriate sector, is replaced by the quasi-local density
\[
\mathcal N^{(F)}(t,\mathbf x)
=
F(\nabla^2/E_M^2)\mathcal N(t,\mathbf x)
=
\int d^3y\,\kappa_F(\mathbf x-\mathbf y)\mathcal N(t,\mathbf y).
\]
So the finite localization width is a consequence of the nonlocal QFT algebra, and is not just a limitation of a particular detector model.} induced by the non-local regulator. The measured density is then the convolution $p_F(x)=(\kappa_F*p)(x),$ and under the stated normalization, centering, and finite-variance assumptions on \(\kappa_F\), we obtain the exact variance-addition law. Combining this identity with the ordinary Heisenberg inequality gives a non-local uncertainty relation. So the minimal localization scale arises from the response width of the non-local observable algebra rather than from a modified canonical commutator, a maximal momentum uncertainty, or a discretization of spacetime. We distinguish ourselves in the the present construction from standard generalized-uncertainty-principle models that have been previously~\cite{KempfManganoMann1995,Kempf1997,Hossenfelder2013,CasadioFengKuntzScardigli2023}.

Recent work has further sharpened the status of these theories like in particular modern analyses have clarified perturbative unitarity and Cutkosky-type rules in nonlocal quantum field theory~\cite{BrisceseModesto2019Cutkosky,KoshelevTokareva2021}, the role of Lorentzian versus Euclidean contour prescriptions~\cite{Buoninfante2022Contour}, the structure of classical and quantum weakly nonlocal gravity~\cite{BasBeneitoCalcagniRachwal2023,BuoninfanteGiacchiniNettoHandbook2023}, functional-integral issues in nonlocal quantum gravity~\cite{Calcagni2024PathIntegral}, anomaly questions in nonlocal gauge-theoretic extensions~\cite{AbuAjamiehChattopadhyayGhoshalOkada2024}, and connections between nonlocal field theory, doubly special relativity, and singularity-free effective structures~\cite{RelancioSantamariaSanz2024,Boos2025Topological,AmelinoCamelia2002DSR,MagueijoSmolin2002,KowalskiGlikman2001DSR,MagueijoSmolin2003,AmelinoCamelia2010DSR,Hossenfelder2007Position,Hossenfelder2007Multiparticle,Hossenfelder2010Locality,Hossenfelder2014SoccerBall,Hossenfelder2015NoGoNetworks}. These developments are important for this paper because the uncertainty relation derived below relies not on a deformed canonical commutator but on the operational localization structure induced by the regulated observable algebra. So this paper should be read in tandem to recent studies of nonlocal unitarity, contour prescriptions, nonlocal gauge consistency, weakly nonlocal gravity, and relativistic localization to make sure that the ideas, results, and frameworks don,t get conflated and confused as we have slightly different approaches that do change the outcomes.

There is as well recent phenomenological applications where I used entire function regulated nonlocal field theory to study heavy-quark threshold dynamics, including the top--antitop threshold enhancement and the effective nonrelativistic kernel
induced by the regulated Bethe--Salpeter equation~\cite{CMSToponium2025,ATLASToponium2026,ThompsonToponium2026}.

\section{The Ordinary Principle of Uncertainty}

We first recall the standard derivation of the ordinary uncertainty relation in a mathematically precise form to lay the foundations that will be used throught the paper.

\begin{definition}
Let $\cH$ be a Hilbert space, let $A$ be a self-adjoint operator on a dense domain $\mathcal{D}(A)\subset \cH$, and let $\psi \in \mathcal{D}(A^2)$ satisfy $\norm{\psi}=1$. The expectation value of $A$ in the state $\psi$ is
\begin{equation}
\langle A \rangle_\psi := \braket{\psi}{A\psi},
\end{equation}
and the variance of $A$ in the state $\psi$ is
\begin{equation}
(\Delta_\psi A)^2 := \braket{\psi}{(A-\langle A\rangle_\psi)^2\psi}.
\end{equation}
We write $\Delta A$ when the state is clear from context.
\end{definition}

\begin{theorem}\cite{Robertson,Schrodinger}[Robertson--Schr\"odinger inequality]
Let $A$ and $B$ be self-adjoint operators on a common invariant dense domain $\mathcal{D}\subset \cH$, and let $\psi \in \mathcal{D}$ with $\norm{\psi}=1$. Then
\begin{equation}
(\Delta A)^2(\Delta B)^2
\ge
\frac{1}{4}\left|\braket{\psi}{[A,B]\psi}\right|^2.
\label{eq:Robertson}
\end{equation}
\end{theorem}

The proof of this is trivial and standard, first set:
\begin{equation}
\widetilde{A}:=A-\langle A\rangle_\psi,
\qquad
\widetilde{B}:=B-\langle B\rangle_\psi,
\end{equation}
then we see:
\begin{equation}
(\Delta A)^2 = \norm{\widetilde{A}\psi}^2,
\qquad
(\Delta B)^2 = \norm{\widetilde{B}\psi}^2.
\end{equation}
and by the Cauchy--Schwarz inequality:
\begin{equation}
\norm{\widetilde{A}\psi}^2 \norm{\widetilde{B}\psi}^2
\ge
\left|\braket{\widetilde{A}\psi}{\widetilde{B}\psi}\right|^2,
\end{equation}
therefore:
\begin{equation}
\braket{\widetilde{A}\psi}{\widetilde{B}\psi}
=
\braket{\psi}{\widetilde{A}\widetilde{B}\psi},
\end{equation}
then we write:
\begin{equation}
\widetilde{A}\widetilde{B}
=
\frac{1}{2}\{\widetilde{A},\widetilde{B}\}
+
\frac{1}{2}[\widetilde{A},\widetilde{B}].
\end{equation}
The expectation value of the anticommutator is real, while the expectation value of the commutator is purely imaginary so therefore:
\begin{equation}
\left|\braket{\psi}{\widetilde{A}\widetilde{B}\psi}\right|^2
\ge
\frac{1}{4}\left|\braket{\psi}{[\widetilde{A},\widetilde{B}]\psi}\right|^2,
\end{equation}
and since constants commute:
\begin{equation}
[\widetilde{A},\widetilde{B}] = [A,B],
\end{equation}
and substituting yields \eqref{eq:Robertson}.

In the one-dimensional Schr\"odinger representation on $\cH=L^2(\R,\dd x)$, one takes:
\begin{equation}
(X\psi)(x)=x\psi(x),\qquad (P\psi)(x)=-i\,\frac{\dd}{\dd x}\psi(x),
\end{equation}
on the standard common domain of smooth rapidly decreasing functions. Then we see:
\begin{equation}
[X,P]=i,
\label{eq:XPcomm}
\end{equation}
so \eqref{eq:Robertson} yields the ordinary Heisenberg relation~\cite{Kennard,Heisenberg}:
\begin{equation}
\Delta X\,\Delta P \ge \frac{1}{2}.
\label{eq:HeisenbergStandard}
\end{equation}

The same argument also holds componentwise in $\R^3$:
\begin{equation}
[X_j,P_k]=i\delta_{jk},
\qquad
\Delta X_j\,\Delta P_j\ge \frac{1}{2},
\end{equation}
for $j,k=1,2,3$.

\section{On Localization in Relativistic Quantum Theory}

In local relativistic quantum theory the interpretation of position $X$ as a fundamental observable becomes representation-dependent and problematic beyond the one-particle sector~\cite{NewtonWigner,Wightman1962,Malament1996,Fleming1965,Miller1972,FoldyWouthuysen}. So we therefore need a new formulation of localization, in the non-local quantum field theory framework we can adopt the detector language to make the outcome apparent~\cite{Busch1999,FewsterVerch,FewsterVerchMeasurement,PapageorgiouFraserMeasurement}.

\begin{definition}
Let $\cH$ be the system Hilbert space, a localization measurement on a time slice is described by a positive operator-valued measure (POVM)
\begin{equation}
E:\mathcal{B}(\R)\to \mathcal{B}(\cH),
\end{equation}
where $\mathcal{B}(\R)$ denotes Borel subsets of $\R$, such that
\begin{equation}
E(R)\ge 0,\qquad E(\R)=\id,
\end{equation}
and for any normalized state $\psi\in \cH$, the probability of detecting the system in the region $R$ is
\begin{equation}
\mu_\psi(R):=\braket{\psi}{E(R)\psi}.
\end{equation}
If the corresponding measure admits a density $p_\psi(x)$, then
\begin{equation}
\mu_\psi(R)=\int_R \dd x\, p_\psi(x),
\end{equation}
and the non-locl position variance is
\begin{align}
(\Delta X_{\mathrm{op}})^2
:&=
\int_{\R} \dd x\, (x-\bar x)^2 p_\psi(x),
\\
\bar x:&=\int_{\R}\dd x\, x\,p_\psi(x).
\end{align}
\end{definition}

In the strictly local case one idealizes the detector as arbitrarily sharp and identifies~\cite{NewtonWigner,Busch1999}:
\begin{equation}
p_\psi(x)=|\psi(x)|^2
\end{equation}
in a Schr\"odinger-type representation, or its appropriate relativistic one-particle analogue. In this idealized local setting we recover the standard uncertainty relation \eqref{eq:HeisenbergStandard}.

The significance of the nonlocal formulation is that it remains meaningful when sharp projectors cease to be physically implementable~\cite{Busch1999,FewsterVerchMeasurement,FalconeContiLocalization,ThompsonOperational}. This is exactly what occurs in the nonlocal theory we work with.

\section{The NonLocal Field Theory}

We now will introduce the non-local framework, throughout $E_M>0$ denotes the nonlocality mass scale and:
\begin{equation}
L_M := E_M^{-1}
\end{equation}
denotes the associated Moffat length scale, recall we work in units $\hbar=c=1$.

\begin{definition}\cite{Moffat1990FiniteNonlocalGauge,EvensMoffatKleppeWoodard1991,MoffatThompsonReg}
[Admissible entire regulator]
Let $F:\C\to\C$ be an entire function. We say that $F$ is an admissible entire regulator if:
\begin{align}
F(0)&=1,\label{eq:F0}\\
F(\bar z)&=\overline{F(z)} \quad \text{for all } z\in\C,\label{eq:Freality}\\
F(z)&\neq 0 \quad \text{for all finite } z\in\C,\label{eq:Fnozeros}
\end{align}
and $F(-p_E^2/E_M^2)$ decays sufficiently rapidly along the Euclidean axis to provide ultraviolet damping.
\end{definition}

Conditions \eqref{eq:F0}--\eqref{eq:Fnozeros} are the standard physical admissibility conditions where the normalization $F(0)=1$ ensures infrared recovery, the reality condition gives hermiticity on real configurations, and the zero-free condition prevents new finite-plane poles and hence unwanted additional propagating degrees of freedom.

A technical point about Lorentzian signature is subtle and important to make is that in the present paper \(F(\Box/E_M^2)\) is not defined by a naive pointwise continuation of the Euclidean heat kernel. The Euclidean expression is used to characterize ultraviolet damping and to motivate the response scale. The Lorentzian smeared observable is instead defined through the covariant functional calculus, or equivalently through a Fourier-multiplier or boundary-value prescription on an appropriate test-function domain~\cite{Efimov1967,AlebastrovEfimov1973,BrisceseModesto2019Cutkosky,KoshelevTokareva2021,Buoninfante2022Contour,Calcagni2024PathIntegral}. In flat space this means:
\begin{equation}
\widetilde{F(\Box/E_M^2)f}(p)
=
F(-p^2/E_M^2)\tilde f(p),
\end{equation}
whenever \(f\) belongs to the chosen domain of definition. The derivation below does not require closing Lorentzian Feynman contours in all complex quadrants. The only kernel property used in the uncertainty derivation is the induced equal-time spatial response profile and its finite second moment.

If $O(x)$ is a local observable of the undeformed theory then the corresponding non-local observable is:
\begin{equation}
O^{(F)}(x):=\left(F(\Box/E_M^2)O\right)(x).
\end{equation}
For a test function $f$ we define the smeared regulated observable:
\begin{equation}
O^{(F)}(f)
:=
\int \dd^4x\, O(x)\,[F(\Box/E_M^2)f](x).
\label{eq:Ofsmear}
\end{equation}
Equivalently, we can write:
\begin{equation}
[F(\Box/E_M^2)f](x)=\int \dd^4y\, K_F(x-y)f(y),
\label{eq:kernelrep}
\end{equation}
so that:
\begin{equation}
O^{(F)}(f)=\int \dd^4x\,\dd^4y\, K_F(x-y)f(y)\,O(x),
\label{eq:Ofkernel}
\end{equation}
where $K_F$ is the convolution kernel obtained from the Fourier transform of the momentum-space form factor.

The crucial structural point is that for a nontrivial entire UV suppressor $K_F$ cannot have compact support so the physical observable algebra is therefore non-local rather than strictly local~\cite{PaleyWiener,MoffatThompsonReg}.

\begin{definition}[Non-local algebra]
Let $O\subset \R^{1,3}$ be an open bounded spacetime region. The non-local algebra associated with the non-local theory is
\begin{equation}
\mathcal{A}_F(O):=
vN\!\left\{
O^{(F)}(f): f\in C_c^\infty(\R^{1,3}),\ \supp(f)\subset O
\right\},
\end{equation}
where $vN(\cdot)$ denotes the von Neumann algebra generated by the indicated set.
\end{definition}

This viewpoint is also consistent with broader algebraic treatments of local observables in gauge-theoretic settings such as the work by V.~Bonzom, M.~Dupuis, F.~Girelli, and Q.~Pan~\cite{Bonzom:2022LocalObs}.

In our framework exact bounded region projectors are not elements of the physical observable algebra. Localization is therefore intrinsically finite-resolution.

\section{The Spatial Kernel at Equal Time}

To derive the uncertainty relation we will now pass from the spacetime kernel $K_F$ to the corresponding equal-time spatial response kernel that determines the position profile seen by a detector.

\begin{definition}[Equal-time spatial response kernel]
Fix a time slice $t=t_0$ and let $\kappa_F:\R\to\R$ be the induced one-dimensional spatial response kernel of the detector on that slice. We assume:
\begin{enumerate}
\item[(i)] $\kappa_F(x)\ge 0$ $\forall$ $x\in\R$;
\item[(ii)] $\displaystyle \int_{\R}\dd x\, \kappa_F(x)=1$;
\item[(iii)] $\displaystyle \int_{\R}\dd x\, x\,\kappa_F(x)=0$;
\item[(iv)] $\displaystyle \sigma_F^2:=\int_{\R}\dd x\, x^2 \kappa_F(x)<\infty$,
\end{enumerate}
\end{definition}

these conditions mean that $\kappa_F$ is a normalized, centered response function with finite second moment. The quantity $\sigma_F^2$ is the variance of the detector broadening induced by the non-local regulator. These assumptions are satisfied by the Gaussian class relevant to the standard entire function regulator.

We let $p(x)$ denote the underlying local localization density of the state on the time slice. The measured, non-local localization density is then:
\begin{equation}
p_F(x):=(\kappa_F * p)(x)
=
\int_{\R}\dd y\, \kappa_F(x-y)\,p(y).
\label{eq:pFconv}
\end{equation}

\begin{definition}[Non-local width]
The non-local width of the state is the standard deviation of the measured density $p_F$:
\begin{align}
(\Delta X_{\mathrm{NL}})^2
:&=
\int_{\R}\dd x\, (x-\bar x_F)^2 p_F(x),
\\
\bar x_F:&=\int_{\R}\dd x\, x\,p_F(x).
\label{eq:DXNLdef}
\end{align}
\end{definition}

\section{On the Broadening of Localization and the Principle of Uncertainty}

We now prove the exact variance-addition law.

\begin{lemma}[Mean of the convoluted density]
Let $p$ be a probability density on $\R$ with finite first moment, and let $\kappa_F$ satisfy the assumptions above. Then the mean of $p_F=\kappa_F*p$ is equal to the mean of $p$:
\begin{equation}
\bar x_F=\bar x,
\qquad
\bar x:=\int_{\R}\dd x\, x\,p(x).
\end{equation}
\end{lemma}

We can prove this by using \eqref{eq:pFconv}:
\begin{align}
\bar x_F
&=
\int_{\R}\dd x\, x\,p_F(x)
=
\int_{\R}\dd x\int_{\R}\dd y\, x\,\kappa_F(x-y)p(y),
\end{align}
then we set $u=x-y$, so $x=u+y$ and $\dd x=\dd u$, then we see:
\begin{align}
\bar x_F
&=
\int_{\R}\dd y\,p(y)\int_{\R}\dd u\, (u+y)\kappa_F(u)\nonumber\\
&=
\int_{\R}\dd y\,p(y)\left(
\int_{\R}\dd u\, u\,\kappa_F(u)
+
y\int_{\R}\dd u\, \kappa_F(u)
\right).
\end{align}
by the centering and normalization conditions of $\kappa_F$:
\begin{equation}
\int_{\R}\dd u\, u\,\kappa_F(u)=0,
\qquad
\int_{\R}\dd u\, \kappa_F(u)=1,
\end{equation}
hence:
\begin{equation}
\bar x_F=\int_{\R}\dd y\, y\,p(y)=\bar x.
\end{equation}

\begin{theorem}[Variance-addition law]
Let $p$ be a probability density on $\R$ with finite second moment, and let $\kappa_F$ satisfy the assumptions above. Then
\begin{equation}
(\Delta X_{\mathrm{NL}})^2
=
(\Delta X)^2+\sigma_F^2,
\label{eq:VarAddition}
\end{equation}
where
\begin{equation}
(\Delta X)^2:=\int_{\R}\dd x\, (x-\bar x)^2p(x)
\end{equation}
is the variance of the underlying local localization density and
\begin{equation}
\sigma_F^2:=\int_{\R}\dd x\, x^2\kappa_F(x)
\end{equation}
is the variance of the response kernel.
\end{theorem}

By the previous lemma $\bar x_F=\bar x$, so therefore:
\begin{align}
(\Delta X_{\mathrm{NL}})^2
&=
\int_{\R}\dd x\, (x-\bar x)^2 p_F(x)\nonumber\\
&=
\int_{\R}\dd x\int_{\R}\dd y\, (x-\bar x)^2 \kappa_F(x-y)p(y),
\end{align}
we again set $u=x-y$, so $x=u+y$:
\begin{align}
(\Delta X_{\mathrm{NL}})^2
=
\int_{\R}\dd y\,p(y)\int_{\R}\dd u\, (u+y-\bar x)^2 \kappa_F(u)\nonumber\\
=
\int_{\R}\dd y\,p(y)\int_{\R}\dd u\,
\bigl[u^2 + 2u(y-\bar x) + (y-\bar x)^2\bigr]\kappa_F(u),
\end{align}
now split the integrals:
\begin{align}
(\Delta X_{\mathrm{NL}})^2
&=
\left(\int_{\R}\dd u\, u^2\kappa_F(u)\right)\left(\int_{\R}\dd y\, p(y)\right)\nonumber\\
&\quad
+
2\left(\int_{\R}\dd u\, u\kappa_F(u)\right)\left(\int_{\R}\dd y\, (y-\bar x)p(y)\right)\nonumber\\
&\quad
+
\left(\int_{\R}\dd u\, \kappa_F(u)\right)\left(\int_{\R}\dd y\, (y-\bar x)^2 p(y)\right).
\end{align}
The first factor is $\sigma_F^2$, the second term vanishes because:
\begin{equation}
\int_{\R}\dd u\, u\kappa_F(u)=0,
\end{equation}
and the third factor is $(\Delta X)^2$. Therefore \eqref{eq:VarAddition} holds.

This theorem is the central structural statement that the nonlocal theory broadens localization by adding the intrinsic kernel variance to the ordinary local variance.

We now combine it with the ordinary uncertainty relation to find:

\begin{theorem}[Uncertainty principle in non-local quantum field theory]
Assume that the underlying local localization width $\Delta X$ and momentum width $\Delta P$ satisfy the ordinary Heisenberg inequality \eqref{eq:HeisenbergStandard}, then the non-local width obeys
\begin{equation}
\Delta X_{\mathrm{NL}}
\ge
\sqrt{\frac{1}{4(\Delta P)^2}+\sigma_F^2}.
\label{eq:NLuncertaintyGeneral}
\end{equation}
\end{theorem}

We see this proof by Theorem 9 \ref{eq:VarAddition}, then using \eqref{eq:HeisenbergStandard}:
\begin{equation}
\Delta X\ge \frac{1}{2\Delta P},
\end{equation}
hence:
\begin{equation}
(\Delta X_{\mathrm{NL}})^2
\ge
\frac{1}{4(\Delta P)^2}+\sigma_F^2.
\end{equation}
Taking square roots then yields \eqref{eq:NLuncertaintyGeneral}.
Equation \eqref{eq:NLuncertaintyGeneral} is the nonlocal uncertainty relation in its cleanest and most general form. It is derived from the actual nonlocal measurement structure of the theory.

\section{The Gaussian Form}

We now specialize to the simplest and most important class of entire function regulators that being namely the Gaussian-type regulator~\cite{Moffat1990FiniteNonlocalGauge,EvensMoffatKleppeWoodard1991,MoffatThompsonReg}:
\begin{equation}
F\!\left(-\frac{p^2}{E_M^2}\right)=\e^{-p^2/E_M^2}
\end{equation}
along the Euclideanized momentum axis.

The exponential or Gaussian form factor is not historically new as exponential nonlocal form factors appear already in early finite radius and nonlocal interaction proposals, and the systematic study of non-localized actions was developed by Pais and Uhlenbeck~\cite{Wataghin1934,PaisUhlenbeck1950}. The same broad Gaussian or entire function idea also appears in later nonlocal field theoretic regularization programs~\cite{Efimov1967,Efimov1974,Moffat1990,EvensMoffatKleppeWoodard1991,KleppeWoodard1992}. Here the Gaussian is used because it is the simplest representative of the admissible entire function class and because it gives an explicit equal time response kernel. The physical claim of this section is then not the novelty of the form factor, but the resulting operational broadening of localization through the induced response kernel.

At equal time the corresponding one-dimensional spatial response kernel is Gaussian. We choose the normalization:
\begin{equation}
\kappa_F(x)
=
\frac{1}{\sqrt{4\pi}\,L_M}\,
\e^{-x^2/(4L_M^2)}.
\label{eq:kappaGaussian}
\end{equation}
We verify directly that:
\begin{equation}
\int_{\R}\dd x\, \kappa_F(x)=1,
\qquad
\int_{\R}\dd x\, x\,\kappa_F(x)=0.
\end{equation}
Its variance is:
\begin{equation}
\sigma_F^2
=
\int_{\R}\dd x\, x^2\kappa_F(x)=2L_M^2=\frac{2}{E_M^2}.
\label{eq:sigmaGaussian}
\end{equation}

\begin{proposition}[Gaussian non-local uncertainty relation]
For the Gaussian response kernel \eqref{eq:kappaGaussian}, the non-local uncertainty relation becomes
\begin{equation}
\Delta X_{\mathrm{NL}}
\ge
\sqrt{\frac{1}{4(\Delta P)^2}+\frac{2}{E_M^2}}.
\label{eq:GaussianNLUR}
\end{equation}
\end{proposition}

This can be proved by substituting \eqref{eq:sigmaGaussian} into \eqref{eq:NLuncertaintyGeneral}.

The precise numerical coefficient in front of $E_M^{-2}$ depends on the convention used in the argument of $F$ and on whether the physically relevant equal-time kernel corresponds to $F$ or $F^2$ in the observable or response sector. The structurally invariant statement is that the correction is of order $E_M^{-2}$ and therefore defines a minimal localization length of order $E_M^{-1}$.

\section{The Infrared Recovery and Minimal Length}

The infrared expansion of \eqref{eq:GaussianNLUR} is obtained by factoring out $(2\Delta P)^{-1}$:
\begin{align}
\Delta X_{\mathrm{NL}}
&\ge
\frac{1}{2\Delta P}
\sqrt{1+\frac{8(\Delta P)^2}{E_M^2}}.
\end{align}
For $\Delta P\ll E_M$, the Taylor expansion:
\begin{equation}
\sqrt{1+\varepsilon}=1+\frac{\varepsilon}{2}+O(\varepsilon^2)
\end{equation}
with $\varepsilon=8(\Delta P)^2/E_M^2$ gives:
\begin{equation}
\Delta X_{\mathrm{NL}}
\ge
\frac{1}{2\Delta P}
+
\frac{2\,\Delta P}{E_M^2}
+
O\!\left(\frac{(\Delta P)^3}{E_M^4}\right).
\label{eq:IRexpansion}
\end{equation}
Thus the usual Heisenberg relation is recovered in the infrared, with a non-local correction suppressed by $E_M^{-2}$. In the ultraviolet:
\begin{equation}
\Delta P\to \infty
\qquad \Longrightarrow \qquad
\Delta X_{\mathrm{NL}}\to \frac{\sqrt{2}}{E_M}.
\label{eq:UVlimit}
\end{equation}
Hence the theory predicts a minimal localization length:
\begin{equation}
\Delta X_{\min}=\frac{\sqrt{2}}{E_M}
\end{equation}
for the Gaussian normalization chosen above.

In Gaussian entire function regulated non-local quantum field theory no physically realizable localization measurement can resolve spatial structure below a length of order $L_M=E_M^{-1}$.

A natural question is whether the existence of a nonzero minimal localization length forces the existence of a maximal momentum uncertainty, it does not\footnote{This question was asked to me by a friend and colleague Hilary Carteret seeing if we could find a way to show the universe has a minimal length scale that is Lorentz invariant and comes from the minimal position uncertainty presented in this paper.}. The point is simple but highly important because the ordinary Heisenberg inequality provides only a lower bound on the product of $\Delta X\,\Delta P$ but not an upper bound on either factor separately. We now will state this precisely and then apply it to the present non-local framework.

\begin{theorem}[Putative theorem: A minimal position uncertainty implies a maximal momentum uncertainty]
Let $x_{\min}>0$ be fixed and suppose that
\begin{align}
\Delta X &\ge x_{\min}, \label{eq:minx_assumption}\\
\notag \Delta X\,\Delta P &\ge \frac{1}{2}.
\end{align}
Then there exists a finite number $P_{\max}>0$ such that
\begin{equation}
\Delta P \le P_{\max}.
\end{equation}
\end{theorem}

To show this proof we shall define an admissible region:
\begin{equation}
\mathcal{R}:=\left\{(x,p)\in (0,\infty)^2 \;:\; x\ge x_{\min}, \; xp\ge \frac{1}{2}\right\}.
\end{equation}
We can show that the projection of $\mathcal{R}$ onto the $p$-axis is unbounded therefore providing a counterexample to the theorem.

First let $p\ge \frac{1}{2x_{\min}}$ and choose:
\begin{equation}
x=x_{\min},
\end{equation}
then $x\ge x_{\min}$ is satisfied, and moreover:
\begin{equation}
xp = x_{\min}p \ge x_{\min}\frac{1}{2x_{\min}}=\frac{1}{2}.
\end{equation}
Hence $(x_{\min},p)\in\mathcal{R}$ for every:
\begin{equation}
p\ge \frac{1}{2x_{\min}}.
\end{equation}
Therefore the projection of $\mathcal{R}$ onto the $p$-axis contains the half-line:
\begin{equation}
\left[\frac{1}{2x_{\min}},\infty\right),
\end{equation}
and so it is unbounded above and thus no finite upper bound on $\Delta P$ follows from \eqref{eq:minx_assumption} and \eqref{eq:HeisenbergStandard}.

This already gives a counterexample in complete generality as the pair:
\begin{equation}
\Delta X=x_{\min}, \qquad \Delta P=p,
\end{equation}
with any $p\ge 1/(2x_{\min})$, satisfies both assumptions while allowing arbitrarily large $\Delta P$. So a minimal position uncertainty by itself does not force a maximal momentum uncertainty and hence the theorem is false by counterexample.

We now apply this observation to the present non-local theory, from Theorem 10 (\eqref{eq:NLuncertaintyGeneral}) we have:
\begin{equation}
\Delta X_{\mathrm{NL}} \ge f(\Delta P),
\qquad
f(p):=\sqrt{\frac{1}{4p^2}+\sigma_F^2},
\label{eq:f_of_p_nonlocal}
\end{equation}
where $p>0$. The derivative is:
\begin{align}
f'(p)
&=
\frac{1}{2}\left(\frac{1}{4p^2}+\sigma_F^2\right)^{-1/2}
\left(-\frac{1}{2p^3}\right)
\\&=
-\frac{1}{4p^3\sqrt{\frac{1}{4p^2}+\sigma_F^2}}<0,
\end{align}
for every $p>0$. Therefore $f(p)$ is strictly decreasing on $(0,\infty)$. It follows that the smallest value allowed by the bound is obtained only in the ultraviolet limit where $p=\Delta P \to \infty,$
and:
\begin{equation}
\inf_{p>0} f(p)=\lim_{p\to\infty} f(p)=\sigma_F.
\end{equation}
So the minimal non-local width is:
\begin{equation}
\Delta X_{\min}=\sigma_F,
\end{equation}
and it is approached asymptotically as $\Delta P\to\infty$, not at any finite maximal momentum uncertainty.

For the Gaussian kernel where $\sigma_F^2=2/E_M^2$ this becomes:
\begin{equation}
\notag\Delta X_{\mathrm{NL}} \ge \sqrt{\frac{1}{4(\Delta P)^2}+\frac{2}{E_M^2}},
\end{equation}
and therefore we see that:
\begin{equation}
\Delta X_{\min}
=
\lim_{\Delta P\to\infty}\Delta X_{\mathrm{NL}}
=
\frac{\sqrt{2}}{E_M}.
\end{equation}
Thus in the present entire function regulated non-local theory the minimal localization length is produced by the finite response width of the non-local observable algebra, not by a maximal momentum uncertainty.

It is also useful to contrast this with the deformed-commutator framework~\cite{KempfManganoMann1995}, we start by noting that there one may obtain a lower bound of the form~\cite{KempfManganoMann1995}:
\begin{equation}
\Delta x \ge \frac{\hbar}{2}\left(\frac{1}{\Delta p}+\beta \Delta p\right),
\end{equation}
whose right-hand side is minimized at the finite value $\Delta p=1/\sqrt{\beta}$. But even there this minimizing value is not a maximal momentum uncertainty as it is only the value at which the lower bound on $\Delta x$ is smallest. Larger values of $\Delta p$ are still allowed, so the same logical distinction must therefore be kept sharply in mind here.

The physical conclusion is that in the present theory the ultraviolet obstruction is not that momentum fluctuations cease to exist beyond some maximal scale but rather it is that arbitrarily large momentum spread no longer yields arbitrarily sharp physically realizable localization. The non-local kernel prevents the operational meaning of pointwise localization below the scale $L_M\sim E_M^{-1}$ even though $\Delta P$ itself remains unbounded.

Although this may seem counterintuitive at first but the logic is actually quite simple once you see it, in the ordinary local theory we tend to think that increasing $\Delta P$ always improves localization because the Heisenberg term $1/(2\Delta P)$ keeps decreasing. If that were the only effect present then yes sharper and sharper localization would indeed be possible. In the present theory however the measured position is not the underlying ideal local position observable but the output of a nonlocal detector response obtained by convolution with the kernel $\kappa_F$. So increasing $\Delta P$ can continue to sharpen the underlying local packet but it cannot remove the finite width of the measuring kernel itself. The situation is therefore analogous to imaging an object with an instrument of finite resolution where making the object narrower does not produce an image narrower than the point spread function of the apparatus. In this sense $\Delta P$ remains unbounded but pointwise localization ceases to be physically realizable below the scale set by the intrinsic nonlocal response width, which is of order $L_M\sim E_M^{-1}$. This is why the ultraviolet obstruction is not a maximal momentum fluctuation, but rather a finite resolution length that is fundamental to our universe.

This should be contrasted with two related but different uses of the words ``uncertainty'' and ``nonlocality.'' In generalized-uncertainty principle models the minimal length is usually introduced kinematically through a deformed commutator or through a modified dispersion relation~\cite{KempfManganoMann1995,Kempf1997,Hossenfelder2013}. In the current framework the canonical uncertainty relation is not deformed at the starting point but the nonlocality comes through the measurement map itself where the physically observed localization density is the output of a quasi-local detector response. Quantum-information analyses can relate uncertainty principles to the strength of non-local correlations~\cite{OppenheimWehner2010}, but that is not the type of nonlocality studied here as the present work concerns spacetime nonlocality in the observable algebra and its effect on operational localization.

There are also detector-based studies of nonlocal or minimal length field theories, for example analyses of modified Wightman functions and Unruh-DeWitt detector responses~\cite{GimUmKim2018,Kajuri2019}. Those works ask how a nonlocal scale modifies detector transition rates, Wightman functions, or thermal responses. The present analysis done here instead isolates the equal-time localization observable and derives the induced uncertainty relation directly from the response kernel.

\section{On the de Broglie Wavelength}

We now will make precise the sense in which the ordinary de Broglie relation survives in a Lorentz-covariant nonlocal quantum field theory while its interpretation as an arbitrarily sharp resolution scale does not\footnote{I would like to give a special acknowledgement to Richard Epp from the University of Waterloo for a very good lecture he gave to the University of Waterloo's quantum club in March of 2026, titled "The Relativistic Backbone of Quantum Physics," where he discussed how de Broglie was able to show a wave particle duality based on Einsteins 1905 paper and how quantum mechanics is rooted in ideas from relativity. This prompted me to explore de Broglie's work and how it could be interpreted in our present framework.}~\cite{deBroglie1924,Einstein1905}. The main point is that special relativity fixes the kinematics of wave propagation through the relativistic dispersion relation but by itself does not impose a minimal localization length, but the minimal length arises only after the physical observable algebra is replaced by nonlocal regulated smearings and the detector response acquires a finite kernel width.

We continue to work in units $\hbar=c=1$, spatial vectors are denoted in boldface, $m\ge 0$ is the particle mass, $E(\mathbf p)=\sqrt{|\mathbf p|^2+m^2}$ is the positive relativistic energy, and $E_M>0$ is the non-locality energy scale with associated Moffat length $L_M := E_M^{-1}$.

We begin with a positive frequency one-particle wave packet in flat spacetime:
\begin{equation}
\Psi(t,\mathbf x)
=
\int_{\mathbb R^3}\frac{d^3p}{(2\pi)^3}\,
a(\mathbf p)\,
e^{-iE(\mathbf p)t+i\mathbf p\cdot\mathbf x},
\label{eq:db_packet}
\end{equation}
where $a(\mathbf p)$ is the momentum-space amplitude normalized so that the state has finite norm, $t$ is the time coordinate, and $\mathbf x\in\mathbb R^3$ is the spatial position vector. If we suppose that $a(\mathbf p)$ is sharply peaked near some momentum $\mathbf p_0\neq 0$ we then expand the relativistic energy around $\mathbf p_0$:
\begin{equation}
E(\mathbf p)
=
E_0
+
\nabla_{\mathbf p}E(\mathbf p)\big|_{\mathbf p=\mathbf p_0}\cdot(\mathbf p-\mathbf p_0)
+
O\!\left(|\mathbf p-\mathbf p_0|^2\right),
\label{eq:energy_expand}
\end{equation}
where:
\begin{equation}
E_0 := E(\mathbf p_0)=\sqrt{|\mathbf p_0|^2+m^2}.
\end{equation}
And since:
\begin{equation}
\nabla_{\mathbf p}E(\mathbf p)=\frac{\mathbf p}{E(\mathbf p)},
\end{equation}
the group velocity of the packet is:
\begin{equation}
\mathbf v_g
=
\nabla_{\mathbf p}E(\mathbf p)\big|_{\mathbf p=\mathbf p_0}
=
\frac{\mathbf p_0}{E_0},
\label{eq:group_velocity}
\end{equation}
equation \eqref{eq:group_velocity} is exactly the ordinary relativistic particle velocity, so the packet envelope propagates in a way fully consistent with special relativity. 

At this point it is useful to make explicit why exactly does arbitrarily large momentum not violate causality as to follow up on the previous section. The reason is that relativistic causality constrains the propagation speed of disturbances not the magnitude of momentum itself. For the dispersion relation:
\[
E(p)=\sqrt{|p|^{2}+m^{2}},
\]
the group velocity is:
\[
v_g=\nabla_p E(p)=\frac{p}{E(p)},
\]
and therefore:
\[
|v_g|=\frac{|p|}{\sqrt{|p|^{2}+m^{2}}}<1
\]
for every finite $|p|$ when $m>0$, while for $m=0$ we have $|v_g|=1$. So taking $|p|$ or $\Delta P$ arbitrarily large shortens the wavelength and increases the spatial frequency content of the packet but it does not allow the packet envelope to propagate outside the light cone. In this sense there is no tension between unbounded momentum and relativistic causality as high momentum changes resolution, not the invariant causal speed limit.

What changes in the present non-local theory is not the relativistic light cone structure itself but the operational meaning of localization. The underlying local packet may contain arbitrarily large momentum components but the measured localization profile is still filtered through the non-local response kernel. Consequently, increasing $\Delta P$ does not generate superluminal propagation and does not force arbitrarily sharp measurable localization. It only sharpens the underlying local profile up to the point where the intrinsic response width of the theory takes over.

The rapidly oscillating phase of the carrier wave is:
\begin{equation}
\varphi(t,\mathbf x)
=
-E_0 t+\mathbf p_0\cdot\mathbf x,
\label{eq:carrier_phase}
\end{equation}
we let:
\begin{equation}
\hat{\mathbf n}:=\frac{\mathbf p_0}{|\mathbf p_0|}
\end{equation}
be the unit vector in the propagation direction, and write $\mathbf x=x_\parallel \hat{\mathbf n}+\mathbf x_\perp$, where $x_\parallel=\hat{\mathbf n}\cdot\mathbf x$ is the longitudinal coordinate. At fixed time $t=t_0$, the phase change under a longitudinal displacement $\Delta x_\parallel$ is:
\begin{equation}
\Delta \varphi
=
|\mathbf p_0|\,\Delta x_\parallel .
\label{eq:phase_shift}
\end{equation}
One full oscillation corresponds to $|\Delta\varphi|=2\pi$ so therefore the de Broglie wavelength is:
\begin{equation}
\lambda_{\mathrm{dB}}
=
\frac{2\pi}{|\mathbf p_0|}.
\label{eq:debroglie_lambda}
\end{equation}
Equation \eqref{eq:debroglie_lambda} is the relativistic de Broglie relation in natural units~\cite{deBroglie1924}. It is perfectly compatible with special relativity because it is derived from the Lorentz-covariant plane-wave phase and the relativistic mass-shell condition. Special relativity constrains the relation between frequency and momentum but it does not itself prevent $|\mathbf p_0|$ from becoming arbitrarily large. So at the purely kinematical level special relativity alone does not imply a minimal spatial wavelength.

It is important to distinguish two different notions, the first being $\lambda_{\mathrm{dB}}$ as the oscillation scale of the carrier wave, second is the actual localization width of the packet is controlled by the spread in momentum, not by the mean momentum alone. In the local theory these are related by the ordinary Heisenberg inequality \eqref{eq:HeisenbergStandard} where $\Delta X$ is the standard deviation of the spatial localization density and $\Delta P$ is the standard deviation of the momentum distribution along the corresponding spatial direction.

So there are two complementary local statements, that a larger mean momentum $|\mathbf p_0|$ makes the carrier wavelength $\lambda_{\mathrm{dB}}$ shorter and a larger momentum spread $\Delta P$ allows a packet to be localized more sharply through \eqref{eq:HeisenbergStandard}.

Both statements point in the same qualitative direction, it tells us that a higher spatial frequency means finer spatial structure. In a strictly local theory this mechanism has no intrinsic lower bound, we should note that this can also be phrased in Fourier language if we let $f(x)$ be a one-dimensional profile then its Fourier representation is:
\begin{equation}
f(x)=\int_{\mathbb R}\frac{dq}{2\pi}\,\widetilde f(q)\,e^{iqx},
\label{eq:fourier_profile}
\end{equation}
where $q$ is the spatial wave number. A profile with characteristic width $a$ requires Fourier support reaching at least to scales $|q|\sim a^{-1}$. Therefore the local theory admits arbitrarily fine resolution in principle because there is no kinematical obstruction to taking $|q|$ arbitrarily large.

\begin{figure}[h]
\centering
\begin{tikzpicture}
\begin{axis}[
width=0.88\columnwidth,
height=0.48\columnwidth,
xmin=-4.5, xmax=4.5,
ymin=0, ymax=1.25,
axis lines=middle,
xtick=\empty,
ytick=\empty,
xlabel={$x$},
ylabel={},
legend style={draw=none, fill=none, font=\footnotesize, at={(0.98,0.98)}, anchor=north east}
]
\addplot[domain=-4:4,samples=300, thick, dashed] {exp(-x^2/0.55)};
\addlegendentry{$p(x)$}
\addplot[domain=-4:4,samples=300, thick, dotted] {0.58*exp(-x^2/3.2)};
\addlegendentry{$\kappa_F(x)$}
\addplot[domain=-4:4,samples=300, thick] {0.74*exp(-x^2/1.85)};
\addlegendentry{$p_F(x)$}
\end{axis}
\end{tikzpicture}
\caption{The underlying local localization density $p(x)$, the equal-time non-local detector response kernel $\kappa_F(x)$, and the measured density $p_F=\kappa_F\ast p$. The non-local theory preserves the state but changes the non-local accessible profile through convolution with a kernel of width of order $L_M=E_M^{-1}$.}
\label{fig:debroglie_convolution}
\end{figure}
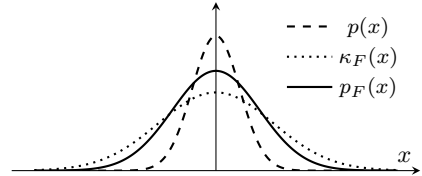

Figure 1 should be read as the operational core of the entire paper, where the underlying local profile $p(x)$ is the localization density one would use in the strictly local theory, while $\kappa_F(x)$ is the equal-time response kernel induced by the regulated observable algebra. The measured profile is not $p(x)$ itself but the convolution $p_F=\kappa_F\ast p$. This is the precise mathematical reason the theory produces a finite localization width since the ultraviolet modification does not first alter the state and then infer a position distribution, but instead alters which position profile is physically accessible to measurement. In particular because $\kappa_F$ is normalized and centered, the mean position is unchanged whereas the variance is enlarged by the fixed kernel contribution $\sigma_F^2$, giving the exact variance-addition law:
\[
(\Delta X_{\mathrm{NL}})^2=(\Delta X)^2+\sigma_F^2.
\]
So Figure 1 is not just illustrative but it is the visual summary of why the non-local theory preserves the infrared notion of localization while replacing exact point resolution by finite operational resolution.

The non-local theory changes this conclusion because the physically measurable localization density is not the underlying local profile $p(x)$ but the convoluted profile:
\begin{equation}
\notag p_F(x)
=
(\kappa_F\ast p)(x)
=
\int_{\mathbb R}dy\,\kappa_F(x-y)p(y),
\label{eq:db_convolution}
\end{equation}
where $\kappa_F$ is the induced equal-time detector response kernel. By assumption we know:
\begin{equation}
\notag\kappa_F(x)\ge 0,
\qquad
\int_{\mathbb R}dx\,\kappa_F(x)=1,
\qquad
\int_{\mathbb R}dx\,x\,\kappa_F(x)=0,
\label{eq:db_kernel_conditions}
\end{equation}
and its variance is:
\begin{equation}
\sigma_F^2
:=
\int_{\mathbb R}dx\,x^2\kappa_F(x).
\label{eq:db_sigmaF}
\end{equation}
The non-local width of the measured profile is then:
\begin{equation}
\notag (\Delta X_{\mathrm{NL}})^2
:=
\int_{\mathbb R}dx\,(x-\bar x_F)^2p_F(x),
\qquad
\bar x_F:=\int_{\mathbb R}dx\,x\,p_F(x).
\label{eq:db_deltaXNL_def}
\end{equation}
We now can state the consequence for de Broglie resolution.

\begin{proposition}[de Broglie limit in Non-local QFT]
Let $\Delta X$ and $\Delta P$ be the underlying local position and momentum widths in one spatial direction, and let $\Delta X_{\mathrm{NL}}$ be the non-locally measured width after non-local smearing by a response kernel of variance $\sigma_F^2$. Then
\begin{equation}
\notag(\Delta X_{\mathrm{NL}})^2=(\Delta X)^2+\sigma_F^2,
\label{eq:db_variance_addition}
\end{equation}
and therefore
\begin{equation}
\notag\Delta X_{\mathrm{NL}}
\ge
\sqrt{\frac{1}{4(\Delta P)^2}+\sigma_F^2}.
\label{eq:db_main_bound}
\end{equation}
In particular,
\begin{equation}
\Delta X_{\mathrm{NL}}\ge \sigma_F.
\label{eq:db_floor_general}
\end{equation}
\end{proposition}

The proof for this can bee seen as \eqref{eq:db_variance_addition} is the exact variance-addition law already established from the convolution structure of the detector response. Combining it with the local Heisenberg inequality \eqref{eq:HeisenbergStandard} of:
\begin{equation}
(\Delta X)^2\ge \frac{1}{4(\Delta P)^2},
\end{equation}
immediately yields \eqref{eq:db_main_bound}. Since the square root on the right-hand side of \eqref{eq:db_main_bound} is bounded below by $\sigma_F$, equation \eqref{eq:db_floor_general} follows.

The mathematical content of \eqref{eq:db_main_bound} is simple but important as it tells us that in a local theory making $\Delta P$ large can drive the lower bound on $\Delta X$ to zero. In the nonlocal theory increasing $\Delta P$ only removes the local part of the bound but it does not remove the kernel contribution $\sigma_F$. So the de Broglie idea that larger momentum gives finer resolution remains true only until the intrinsic response width of the theory is reached.

To make this apparent we will look at the Gaussian regulator where we have:
\begin{equation}
\notag\kappa_F(x)
=
\frac{1}{\sqrt{4\pi}\,L_M}\,
e^{-x^2/(4L_M^2)},
\label{eq:db_gaussian_kernel}
\end{equation}
and hence as before:
\begin{equation}
\notag\sigma_F^2=2L_M^2=\frac{2}{E_M^2},
\label{eq:db_gaussian_variance}
\end{equation}
then substituting \eqref{eq:db_gaussian_variance} into \eqref{eq:db_main_bound} gives us:
\begin{equation}
\notag\Delta X_{\mathrm{NL}}
\ge
\sqrt{\frac{1}{4(\Delta P)^2}+\frac{2}{E_M^2}}.
\label{eq:db_gaussian_bound}
\end{equation}
Taking the ultraviolet limit $\Delta P\to\infty$ yieldsL
\begin{equation}
\Delta X_{\mathrm{NL}}
\longrightarrow
\frac{\sqrt{2}}{E_M}
=
\sqrt{2}\,L_M.
\label{eq:db_gaussian_floor}
\end{equation}

Equation \eqref{eq:db_gaussian_floor} is the precise sense in which the nonlocal theory modifies the de Broglie microscope. This is the familiar microscope intuition of standard expositions of quantum theory except that here the operational resolution saturates at the non-local scale rather than improving without bound \cite{Feynman1965Character,FeynmanLeightonSands}. The theory does not invalidate the relativistic de Broglie relation \eqref{eq:debroglie_lambda} but rather it changes the interpretation of what can be operationally resolved, meaning decreasing the wavelength indefinitely does not force the measured localization width to zero. Instead the observed width saturates at a nonzero value of order $L_M$.

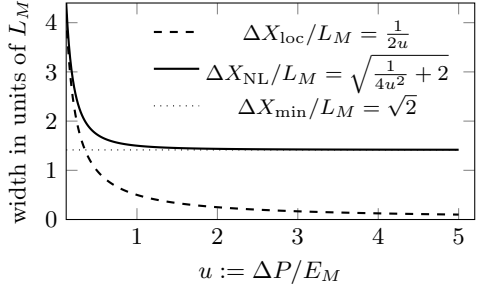
\begin{figure}[h]
\centering
\begin{tikzpicture}
\begin{axis}[
width=0.88\columnwidth,
height=0.56\columnwidth,
xmin=0.12, xmax=5.2,
ymin=0, ymax=4.4,
xlabel={$u:=\Delta P/E_M$},
ylabel={width in units of $L_M$},
legend style={draw=none, fill=none, font=\footnotesize, at={(0.97,0.97)}, anchor=north east}
]
\addplot[domain=0.12:5,samples=300, thick, dashed] {1/(2*x)};
\addlegendentry{$\Delta X_{\mathrm{loc}}/L_M=\frac{1}{2u}$}
\addplot[domain=0.12:5,samples=300, thick] {sqrt(1/(4*x^2)+2)};
\addlegendentry{$\Delta X_{\mathrm{NL}}/L_M=\sqrt{\frac{1}{4u^2}+2}$}
\addplot[domain=0.12:5,samples=2, dotted] {sqrt(2)};
\addlegendentry{$\Delta X_{\min}/L_M=\sqrt{2}$}
\end{axis}
\end{tikzpicture}
\caption{Comparison between the local uncertainty bound and the Gaussian non-local bound. In the local theory the lower bound decreases like $(2\Delta P)^{-1}$, whereas in the non-local theory it saturates at the minimal non-local width $\Delta X_{\min}=\sqrt{2}\,L_M$.}
\label{fig:debroglie_saturation}
\end{figure}

Figure 2 makes visible three distinct momentum-space regimes, the first is in the infrared regime $\Delta P \ll E_M$ where the non-local curve lies very close to the ordinary local Heisenberg curve, so the theory is effectively indistinguishable from the local one. Second at intermediate momentum spread $\Delta P \sim E_M$, the non-local correction becomes comparable to the local term and the two curves separate appreciably. The third is in the ultraviolet regime $\Delta P \gg E_M$ where the local contribution $1/(2\Delta P)$ continues to fall, but the measurable width no longer follows it and instead saturates at the constant value $\Delta X_{\min}=\sqrt{2}\,L_M$. In this way Figure 2 displays the exact point at which the de Broglie intuition changes interpretation, that larger momentum spread still sharpens the underlying local packet but it no longer sharpens the physically realizable localization profile after the intrinsic response scale of the theory is reached. The figure therefore shows directly that the ultraviolet obstruction is a finite resolution floor, not a maximal momentum uncertainty.

From an experimental point of view, Figure 2 is the cleanest diagnostic of the theory as a localization experiment with tunable momentum spread would distinguish the local and non-local frameworks by testing whether the observed width continues to fall like $(2\Delta P)^{-1}$ or instead saturates at a nonzero value of order $L_M$.

Now the final idea of this section is that special relativity determines the covariant kinematics of the wave packet through the relativistic mass shell and the Lorentz-invariant phase. De Broglie's relation then identifies the spatial oscillation scale associated with momentum. In a strictly local theory this shorter wavelength can in principle be converted into arbitrarily fine resolution. In the present nonlocal theory this final step fails but not because Lorentz covariance is broken but because the physical observable algebra is nonlocal and every localization measurement is filtered by a response kernel of nonzero width.

This is exactlly the content of the covariance theorem proved in~\cite{ThompsonCovariance}, the entire function deformation modifies the localization sector while leaving the Lorentz-covariant kinematics intact.

So the correct statement in nonlocal quantum field theory is not that de Broglie's wavelength ceases to exist but reallt the correct statement is much sharper:
\begin{equation}
\lambda_{\mathrm{dB}}\downarrow 0
\quad \not\!\!\Longrightarrow \quad
\Delta X_{\mathrm{NL}}\downarrow 0.
\label{eq:db_not_implies}
\end{equation}
Instead we see that:
\begin{equation}
\Delta X_{\mathrm{NL}}
\ge
O(L_M).
\label{eq:db_operational_floor}
\end{equation}
So the ultraviolet theory remains Lorentz-covariant but pointwise localization is no longer a physically realizable observable notion below the Moffat length.

We have learned as well that de Broglie’s insight survives but its interpretation changes in the UV, so matter still has the relativistic wave character, momentum still sets a wavelength, and special relativity still fixes the dispersion relation. What changes is that shorter wavelength no longer implies arbitrarily precise localization of a physical event. Below $L_M$ the theory says nature will only let you access smeared finite-resolution observables. So de Broglie remains right, but in the UV his wavelength becomes a probe scale inside a quasi-local observable algebra rather than a route to exact point resolution.

\section{On Measurement and Commutators}

A common question is whether \eqref{eq:GaussianNLUR} should be interpreted as coming from a modified canonical commutator~\cite{KempfManganoMann1995}? Now the answer is that the derivation given here does not require such a postulate.

The logic is the nonlocal theory replaces strictly local observables by nonlocal regulated smearings, then the corresponding detector response to localization is broadened by the non-local kernel, this broadening gives the exact variance-addition law, and combining variance addition with the local Heisenberg inequality yields the non-local uncertainty relation.

Thus the new uncertainty relation is derived from the nonlocal observable algebra and the measurement structure of the theory itself~\cite{FewsterVerch,FewsterVerchMeasurement}.

Questions of operator ordering and the status of Wick ordering in nonlocal UV completions are separate from the present derivation but they reinforce the general point that the ultraviolet modification is implemented at the level of regulated observables rather than by postulating a deformed canonical algebra~\cite{T:Wick}.

One may if desired introduce an effective commutator whose Robertson bound reproduces the leading infrared expansion \eqref{eq:IRexpansion}. For example we can define formally:
\begin{equation}
[X,P]_{\mathrm{eff}}=i\left(1+\beta P^2+\cdots\right)
\end{equation}
with $\beta$ chosen so that the induced lower bound matches the first correction term. But this effective commutator is not the primary derivation as it is only a convenient reparametrization of the already derived non-local bound.

\section{On Conformal Invariance and Maxwell Theory}

It would be useful to distinguish two conceptually different ways in which a minimal length may enter quantum theory to help understand the underlying motivation, the first way we should touch on is the deformed canonical algebra approach associated with Kempf, Mangano, and Mann where in that framework they modify the Heisenberg algebra itself so that the existence of a nonzero minimal position uncertainty is built into the kinematics from the start~\cite{KempfManganoMann1995}. The second point is our entire function regulated nonlocal approach used in the present work where we do not deform the canonical commutator but instead we replace strictly local observables by quasi-local regulated observables built from an entire function of the d'Alembertian. The two frameworks both introduce an intrinsic ultraviolet scale but they do so in very different ways and this difference matters for conformal symmetry and for Maxwell theory\footnote{I would like to thank Robert Mann for discussing the 1996 paper with me and pointing out the violation of conformal invariance to me as that prompted me to write this section and I believe it improved the paper a great deal.}.

In the simplest one-dimensional version of the minimal length framework one replaces the ordinary Heisenberg algebra:
\begin{equation}
[x,p]=i\hbar
\end{equation}
by the deformed commutator:
\begin{equation}
[x,p]=i\hbar\bigl(1+\beta p^{2}\bigr),
\label{eq:KMMcomm}
\end{equation}
where $\beta>0$ is a new constant with dimensions of inverse momentum squared, though equivalently $\sqrt{\beta}\,\hbar$ has dimensions of length and therefore $\beta$ introduces an intrinsic microscopic scale into the theory.

Starting from \eqref{eq:KMMcomm}, the Robertson inequality gives us:
\begin{equation}
\Delta x\,\Delta p \geq \frac{1}{2}\left|\langle [x,p]\rangle\right|
= \frac{\hbar}{2}\left(1+\beta \langle p^{2}\rangle\right),
\end{equation}
then using:
\begin{equation}
\langle p^{2}\rangle=(\Delta p)^{2}+\langle p\rangle^{2},
\end{equation}
we obtain:
\begin{equation}
\Delta x\,\Delta p \geq \frac{\hbar}{2}\left(1+\beta (\Delta p)^{2}+\beta \langle p\rangle^{2}\right).
\label{eq:KMMunc0}
\end{equation}
For the states that minimize the position uncertainty we may take $\langle p\rangle=0$ which reduces \eqref{eq:KMMunc0} to:
\begin{equation}
\notag\Delta x \geq \frac{\hbar}{2}\left(\frac{1}{\Delta p}+\beta \Delta p\right).
\label{eq:KMMunc1}
\end{equation}
To find the minimum we differentiate the right-hand side with respect to $\Delta p$:
\begin{equation}
\frac{d}{d(\Delta p)}
\left[
\frac{\hbar}{2}\left(\frac{1}{\Delta p}+\beta \Delta p\right)
\right]
=
\frac{\hbar}{2}
\left(
-\frac{1}{(\Delta p)^2}+\beta
\right),
\end{equation}
then setting this equal to zero gives us:
\begin{equation}
\Delta p = \frac{1}{\sqrt{\beta}},
\end{equation}
and substituting back into \eqref{eq:KMMunc1} will yeild:
\begin{equation}
\Delta x_{\min}=\hbar \sqrt{\beta}.
\label{eq:KMMmin}
\end{equation}
So the scale $\beta$ is not an emergent consequence of dynamics but it is built directly into the algebra from the start.

Now we note that this breaks exact scale and conformal invariance as classical conformal invariance contains dilatations as a subgroup, so it is enough to examine scale transformations. Under a dilatation:
\begin{equation}
x \mapsto x'=\lambda x,
\qquad
p \mapsto p'=\lambda^{-1}p,
\end{equation}
the transformed commutator is:
\begin{equation}
[x',p']=[\lambda x,\lambda^{-1}p]=[x,p]
=i\hbar\bigl(1+\beta p^{2}\bigr),
\end{equation}
and since $p=\lambda p'$ this becomes:
\begin{equation}
[x',p']=i\hbar\bigl(1+\beta \lambda^{2} p'^{2}\bigr).
\label{eq:scaledKMM}
\end{equation}
If the theory were exactly scale invariant then \eqref{eq:scaledKMM} would have to retain the same functional form with the same fixed parameter $\beta$:
\begin{equation}
[x',p']\stackrel{?}{=} i\hbar\bigl(1+\beta p'^{2}\bigr),
\end{equation}
but this is impossible unless either $\lambda=1$ or $\beta$ is allowed to transform. But we note that $\beta$ is supposed to be a fixed parameter of the theory so therefore the deformation \eqref{eq:KMMcomm} is not invariant under dilatations, and since conformal invariance contains dilatations exact conformal invariance is broken as well. This means the breaking is easy to identify since it is caused by the appearance of a fixed new scale $\beta$ in the fundamental commutator.

In four spacetime dimensions the ordinary source-free Maxwell theory:
\begin{equation}
\partial_{\mu}F^{\mu\nu}=0,
\qquad
\partial_{[\mu}F_{\nu\rho]}=0,
\label{eq:Maxwelllocal}
\end{equation}
with action of the form:
\begin{equation}
S_{\text{Max}}=-\frac{1}{4}\int d^{4}x\,F_{\mu\nu}F^{\mu\nu},
\label{eq:MaxwellActionLocal}
\end{equation}
is classically scale invariant and in fact conformally invariant~\cite{Bateman1909,Cunningham1910,ElShowkNakayamaRychkov2011}, the reason is that in $d=4$ the field strength has engineering dimension two, so under a Weyl transformation of some gauge field:
\begin{equation}
x\mapsto \lambda x,
\qquad
A_{\mu}(x)\mapsto \lambda^{-1}A'_{\mu}(x'),
\end{equation}
we have:
\begin{equation}
F_{\mu\nu}(x)\mapsto \lambda^{-2}F'_{\mu\nu}(x'),
\end{equation}
and thus:
\begin{equation}
d^{4}x\,F_{\mu\nu}F^{\mu\nu}
\mapsto
\lambda^{4}\,d^{4}x'\,\lambda^{-4}F'_{\mu\nu}F'^{\mu\nu}
=
d^{4}x'\,F'_{\mu\nu}F'^{\mu\nu}
\end{equation}
and hence $S_{\text{Max}}$ is invariant.

Equivalently in four dimensions the stress tensor may be improved so that the conformal structure is made manifest in the sense analyzed by Callan, Coleman, and Jackiw~\cite{CallanColemanJackiw1970}.

We note that the connection between scale/conformal symmetry and conserved currents is classical~\cite{Noether1918,BesselHagen1921,BanadosReyes2016}.

Once a fixed microscopic length scale is introduced that scale necessarily spoils this exact symmetry but in a minimal length realization of electrodynamics we should expect the field equations or the action to depend on the new parameter $\beta$ through higher-derivative or nonlocal structures. Schematically we are led to an equation of the form:
\begin{equation}
\mathcal{K}(\beta \Box)\,\partial_{\mu}F^{\mu\nu}=0,
\label{eq:schematicKMMMaxwell}
\end{equation}
or to an equivalent modification of the action. The precise operator $\mathcal{K}$ is model-dependent but the structural point is not, since the dimensionless argument is $\beta \Box$, and under a dilatation:
\begin{equation}
\Box \mapsto \lambda^{-2}\Box',
\end{equation}
so:
\begin{equation}
\beta \Box \mapsto \beta \lambda^{-2}\Box',
\end{equation}
and so the modified equation cannot remain strictly invariant unless $\beta$ also rescales. To say it again is because $\beta$ is a fixed physical parameter then exact scale and conformal invariance are lost.

So in the minimal length framework the breaking of conformal invariance is not accidental but it is a direct consequence of putting a fixed microscopic scale into the kinematics.

Now the framework used in this paper is different since at the starting point as we do not modify the canonical commutator but instead if $O(x)$ is a local observable of the undeformed theory we define the corresponding regulated observable by~\cite{Moffat1990FiniteNonlocalGauge,EvensMoffatKleppeWoodard1991,MoffatThompsonReg}:
\begin{equation}
\notag O^{(F)}(x)=\bigl(F(\Box/E_{M}^{2})\,O\bigr)(x),
\label{eq:OFdef}
\end{equation}
where $F$ is an admissible entire function and $E_{M}$ is the nonlocality scale. Equivalently recall how we write in the smeared form:
\begin{equation}
\notag O^{(F)}(f)=\int d^{4}x\,O(x)\,[F(\Box/E_{M}^{2})f](x),
\label{eq:OFsmeared}
\end{equation}
the localization width is then obtained from the actual detector response induced by the regulated observable algebra not from a postulated deformation of $[x,p]$.

In particular since we now know the nonlocal uncertainty relation has the form:
\begin{equation}
\notag \Delta X_{\text{NL}} \geq
\sqrt{\frac{1}{4(\Delta P)^{2}}+\sigma_{F}^{2}},
\label{eq:ourNLunc}
\end{equation}
where $\sigma_{F}^{2}$ is the variance of the induced equal-time response kernel. For the Gaussian regulator we have:
\begin{equation}
F\!\left(-\frac{p_{E}^{2}}{E_{M}^{2}}\right)=e^{-p_{E}^{2}/E_{M}^{2}},
\qquad
\sigma_{F}^{2}=\frac{2}{E_{M}^{2}},
\end{equation}
so that as before:
\begin{equation}
\notag \Delta X_{\text{NL}} \geq
\sqrt{\frac{1}{4(\Delta P)^{2}}+\frac{2}{E_{M}^{2}}},
\label{eq:GaussianNLunc}
\end{equation}
the key point to say is that the correction arises from the finite detector response width generated by the regulated observable algebra and it is therefore a statement about localization not a deformation of the basic phase-space commutator.

Even though our framework is quite different from the deformed commutator approach it still contains a fixed ultraviolet scale denoted by $E_{M}$ so we note that in this the exact scale invariance is as well broken, and indeed under a dilatation:
\begin{equation}
x\mapsto x'=\lambda x,
\qquad
\partial_{\mu}\mapsto \lambda^{-1}\partial'_{\mu},
\qquad
\Box \mapsto \lambda^{-2}\Box',
\end{equation}
the argument of the regulator transforms as:
\begin{equation}
\frac{\Box}{E_M^{2}}
\mapsto
\frac{\lambda^{-2}\Box'}{E_M^{2}},
\label{eq:scalebox}
\end{equation}
and therefore:
\begin{equation}
F\!\left(\frac{\Box}{E_M^{2}}\right)
\mapsto
F\!\left(\frac{\lambda^{-2}\Box'}{E_M^{2}}\right),
\end{equation}
which is not equal to $F(\Box'/E_M^{2})$ unless either $\lambda=1$ or $E_M$ is allowed to transform. Since $E_M$ is a fixed physical scale then an exact dilatation invariance is absent and that means exact conformal invariance is absent as well\footnote{We should note that if the universe admits a dynamical nonlocality scale $E_M$ then we may restore conformal invariance.}.

So we do not in fact preserve exact conformal invariance but the distinction comes from the fact that in these minimal-length frameworks the point is not whether a scale is present, but it is how the scale enters the theory. In our case it enters through a Lorentz and gauge covariant function of the d'Alembertian not through a deformation of the canonical algebra.

So then the question is why is Maxwell theory is not spoiled in the same way? Well the essential reason is that the regulator is implemented covariantly. In the gauge sector the relevant operator is not the ordinary box but the gauge-covariant d'Alembertian or Laplace--Beltrami operator:
\begin{equation}
\Box_{\mathrm{cov}}=g^{\mu\nu}D_{\mu}D_{\nu}.
\label{eq:covbox}
\end{equation}
Accordingly a regulated Abelian gauge action in flat spacetime takes the form~\cite{Moffat1990FiniteNonlocalGauge,EvensMoffatKleppeWoodard1991}
:
\begin{equation}
S_{F}[A]=-\frac{1}{4}\int d^{4}x\,
F_{\mu\nu}\,\mathcal{F}\!\left(\frac{\Box}{E_M^{2}}\right)F^{\mu\nu},
\label{eq:RegMaxwellAction}
\end{equation}
where $\mathcal{F}$ is an admissible entire function with $\mathcal{F}(0)=1$. We now vary the action, since:
\begin{equation}
\delta F_{\mu\nu}=\partial_{\mu}\delta A_{\nu}-\partial_{\nu}\delta A_{\mu},
\end{equation}
we obtain:
\begin{align}
\delta S_{F}
&=
-\frac{1}{2}\int d^{4}x\,
\delta F_{\mu\nu}\,
\mathcal{F}\!\left(\frac{\Box}{E_M^{2}}\right)F^{\mu\nu}
\\
&=
-\int d^{4}x\,
(\partial_{\mu}\delta A_{\nu})\,
\mathcal{F}\!\left(\frac{\Box}{E_M^{2}}\right)F^{\mu\nu},
\end{align}
where antisymmetry of $F^{\mu\nu}$ was used in the second step, then integrating by parts gives us:
\begin{equation}
\delta S_{F}
=
\int d^{4}x\,
\delta A_{\nu}\,
\partial_{\mu}
\left[
\mathcal{F}\!\left(\frac{\Box}{E_M^{2}}\right)F^{\mu\nu}
\right].
\end{equation}
And therefore the Euler--Lagrange equations are:
\begin{equation}
\partial_{\mu}
\left[
\mathcal{F}\!\left(\frac{\Box}{E_M^{2}}\right)F^{\mu\nu}
\right]
=0.
\label{eq:RegMaxwellEq}
\end{equation}
In flat Abelian theory $[\partial_{\mu},\Box]=0$ so this can also be written as:
\begin{equation}
\mathcal{F}\!\left(\frac{\Box}{E_M^{2}}\right)\partial_{\mu}F^{\mu\nu}=0.
\label{eq:RegMaxwellEq2}
\end{equation}
The Bianchi identity is unchanged:
\begin{equation}
\partial_{[\mu}F_{\nu\rho]}=0.
\label{eq:BianchiUnchanged}
\end{equation}
Several facts are now immediate to see, the first is that gauge invariance is preserved because the action depends on $A_{\mu}$ only through $F_{\mu\nu}$ and in the non-Abelian generalization through the covariant operator built from $D_{\mu}$. The second is that Lorentz covariance is preserved because the regulator is a scalar function of the covariant d'Alembertian rather than a noncovariant spatial cutoff. Third is if the infrared theory reduces to ordinary Maxwell electrodynamics. For soft momenta we have:
\begin{equation}
\mathcal{F}\!\left(-\frac{p^{2}}{E_M^{2}}\right)
=
1+\mathcal{O}\!\left(\frac{p^{2}}{E_M^{2}}\right),
\qquad
p^{2}\ll E_M^{2},
\end{equation}
and therefore \eqref{eq:RegMaxwellEq2} reduces to:
\begin{equation}
\partial_{\mu}F^{\mu\nu}=0
\end{equation}
up to corrections suppressed by powers of $E_M^{-2}$.

This is a special case of the more general covariance theorem for entire function deformations proved in~\cite{ThompsonCovariance}.

To make the gauge argument and the photon spectrum completely explicit as I have previously had interactions where people have not understood the argument being made that the theory is gauge invarient and the photon remains massless\footnote{A paper written by myself and John Moffat should have derived this but it has now been published in Annalen der Physik as of April 2026 so it will be published here instead~\cite{MoffatThompsonReg}.}~\cite{Schwinger1951,Schwinger1962a,Schwinger1962b,Ward1950,Takahashi1957,MT:ReplyToCline}. We first let:
\begin{equation}
S[A] = -\frac{1}{4}\int d^4x \, F_{\mu\nu}\,\mathcal{F}\!\left(\frac{\Box}{E_M^2}\right) F^{\mu\nu},
\end{equation}
where $F_{\mu\nu} := \partial_\mu A_\nu - \partial_\nu A_\mu$, and where $\mathcal{F}(z)$ is an admissible entire function regulator where the first condition ensures infrared recovery, while the second ensures that no new finite-plane zeros are introduced into the free kinetic operator.

We now want to prove two statements, first that the regulated Abelian theory remains exactly gauge invariant, and that the photon remains massless.

We begin with gauge invariance, so under the Abelian gauge transformation:
\begin{equation}
A_\mu(x)\longrightarrow A'_\mu(x)=A_\mu(x)+\partial_\mu \alpha(x),
\end{equation}
the field strength transforms as:
\begin{align}
F'_{\mu\nu}
&= \partial_\mu A'_\nu - \partial_\nu A'_\mu \\
&= \partial_\mu(A_\nu+\partial_\nu\alpha)-\partial_\nu(A_\mu+\partial_\mu\alpha) \\
&= \partial_\mu A_\nu - \partial_\nu A_\mu
   + \partial_\mu\partial_\nu\alpha-\partial_\nu\partial_\mu\alpha \\
&= F_{\mu\nu},
\end{align}
because partial derivatives commute:
\begin{equation}
\partial_\mu\partial_\nu\alpha=\partial_\nu\partial_\mu\alpha.
\end{equation}
Therefore the action is exactly invariant:
\begin{align}
S[A']&=
-\frac{1}{4}\int d^4x \, F'_{\mu\nu}\,\mathcal{F}\!\left(\frac{\Box}{E_M^2}\right) F'^{\mu\nu}
\\&=
-\frac{1}{4}\int d^4x \, F_{\mu\nu}\,\mathcal{F}\!\left(\frac{\Box}{E_M^2}\right) F^{\mu\nu}
= S[A].
\end{align}

This already excludes an explicit Proca mass term~\cite{Proca1936}, and indeed if we were to add:
\begin{equation}
S_{\text{mass}}=\frac{m_\gamma^2}{2}\int d^4x \, A_\mu A^\mu,
\end{equation}
then under $A_\mu\to A_\mu+\partial_\mu\alpha$ we would obtain:
\begin{align}
\delta_\alpha S_{\text{mass}}
&=
\frac{m_\gamma^2}{2}\int d^4x \,
\Big[(A_\mu+\partial_\mu\alpha)(A^\mu+\partial^\mu\alpha)-A_\mu A^\mu\Big] \\
&=
m_\gamma^2\int d^4x\, A^\mu\partial_\mu\alpha
+\frac{m_\gamma^2}{2}\int d^4x\, \partial_\mu\alpha\,\partial^\mu\alpha,
\end{align}
which does not vanish for a general gauge parameter $\alpha$ and hence an explicit photon mass term is incompatible with the exact gauge symmetry of the regulated Abelian action. By contrast a Stueckelberg completion provides us a gauge-invariant description of a massive Abelian vector field but that is not the structure realized in the present entire function regulated theory~\cite{Stueckelberg1938,RueggRuizAltaba2004}.

We now will derive the free field equations and the corresponding dispersion relation, so we start by varying the action, and this gives us:
\begin{equation}
\notag \partial_\mu\!\left[\mathcal{F}\!\left(\frac{\Box}{E_M^2}\right)F^{\mu\nu}\right]=0.
\end{equation}
In flat Abelian theory one has:
\begin{equation}
[\partial_\mu,\Box]=0,
\end{equation}
so the equations can also be written as:
\begin{equation}
\notag\mathcal{F}\!\left(\frac{\Box}{E_M^2}\right)\partial_\mu F^{\mu\nu}=0,
\end{equation}
and using:
\begin{equation}
\partial_\mu F^{\mu\nu}
=
\partial_\mu(\partial^\mu A^\nu-\partial^\nu A^\mu)
=
\Box A^\nu-\partial^\nu(\partial\cdot A),
\end{equation}
the free equations become:
\begin{equation}
\mathcal{F}\!\left(\frac{\Box}{E_M^2}\right)
\Big[\Box A^\nu-\partial^\nu(\partial\cdot A)\Big]=0.
\end{equation}

To analyze the particle content we pass to momentum space and write:
\begin{equation}
A^\mu(x)=\varepsilon^\mu(p)e^{-ip\cdot x},
\end{equation}
so that:
\begin{align}
\Box A^\mu(x) &= -p^2 A^\mu(x),
\\
\partial^\nu(\partial\cdot A)&= -p^\nu (p\cdot \varepsilon)e^{-ip\cdot x}.
\end{align}
Substituting into the field equation gives us:
\begin{equation}
\mathcal{F}\!\left(-\frac{p^2}{E_M^2}\right)
\Big[-p^2\varepsilon^\nu + p^\nu(p\cdot\varepsilon)\Big]=0.
\end{equation}
For a physical photon polarization we impose transversality:
\begin{equation}
p\cdot\varepsilon=0,
\end{equation}
and then we see the equation reduces to:
\begin{equation}
\mathcal{F}\!\left(-\frac{p^2}{E_M^2}\right)p^2\varepsilon^\nu=0.
\end{equation}
Now $\varepsilon^\nu\neq 0$ for a nontrivial mode, and by assumption:
\begin{equation}
\mathcal{F}(z)\neq 0
\qquad \forall \;\;\text{Finite}\;\; z,
\end{equation}
so the only way this equation can hold is:
\begin{equation}
p^2=0.
\end{equation}
This is exactly the massless dispersion relation, so the propagating Abelian gauge boson in
the regulated theory is still a massless photon. This is also consistent with the very strong experimental upper bounds on any nonzero photon mass~\cite{GoldhaberNieto2010}.

The same conclusion is visible directly from the quadratic kinetic operator as integrating by parts we see that the free action can be written as:
\begin{equation}
S^{(2)}[A]
=
\frac{1}{2}\int d^4x \,
A_\mu
\Big(\eta^{\mu\nu}\Box-\partial^\mu\partial^\nu\Big)
\mathcal{F}\!\left(\frac{\Box}{E_M^2}\right)
A_\nu .
\end{equation}
In momentum space this becomes:
\begin{equation}
S^{(2)}[A]
=
\frac{1}{2}\int \frac{d^4p}{(2\pi)^4}\,
\widetilde{A}_\mu(-p)\,
K^{\mu\nu}(p)\,
\widetilde{A}_\nu(p),
\end{equation}
with:
\begin{equation}
K^{\mu\nu}(p)
=
\mathcal{F}\!\left(-\frac{p^2}{E_M^2}\right)
\Big(-p^2\eta^{\mu\nu}+p^\mu p^\nu\Big).
\end{equation}
This operator is transverse, so:
\begin{equation}
p_\mu K^{\mu\nu}(p)=0.
\end{equation}
That is the momentum-space statement of the Abelian Ward identity. The regulator modifies the
overall analytic weight of the kinetic operator but it does not spoil its transverse gauge structure.

If one now adds the standard covariant gauge-fixing term:
\begin{equation}
S_{\mathrm{gf}}
=
-\frac{1}{2\xi}\int d^4x \,
(\partial\cdot A)\,
\mathcal{F}\!\left(\frac{\Box}{E_M^2}\right)
(\partial\cdot A),
\end{equation}
the quadratic operator becomes invertible and the propagator takes the form:
\begin{equation}
D_{\mu\nu}(p)
=
\frac{-i}{\mathcal{F}\!\left(-\frac{p^2}{E_M^2}\right)}
\frac{1}{p^2+i0}
\left(
\eta_{\mu\nu}
-(1-\xi)\frac{p_\mu p_\nu}{p^2}
\right).
\end{equation}
Because $\mathcal{F}$ is entire and has no finite zeros, it introduces no additional poles. The
only physical pole is the usual one at $p^2=0$, so the photon remains massless and no extra Abelian gauge-boson degrees of freedom appear.

The final conclusion is that in the entire function regulated Abelian gauge sector exact conformal invariance is lost because the theory contains the fixed ultraviolet scale $E_M$, but gauge invariance is preserved exactly and the propagating photon remains massless. The regulator changes the ultraviolet behavior of the theory without changing the infrared gauge
content.

It is worth mentioning that in the earlier nonlocal regularization scheme of Evens, Moffat, Kleppe, and Woodard, gauge consistency of the one-loop vacuum polarization was recovered only after summing three distinct contributions, from the diagram with two cubic nonlocal vertices, the diagram with a quartic nonlocal vertex, and an additional contribution coming from
the functional measure. In that formulation the first two graph classes alone are not transverse as the measure contribution is essential to restore the Ward identity and preserve the masslessness of the photon at one loop. But in the present entire function framework the Abelian gauge sector is written from the outset in a manifestly gauge-covariant form so the free kinetic operator is already transverse and the massless photon pole follows directly from the zero-free property of the regulator.

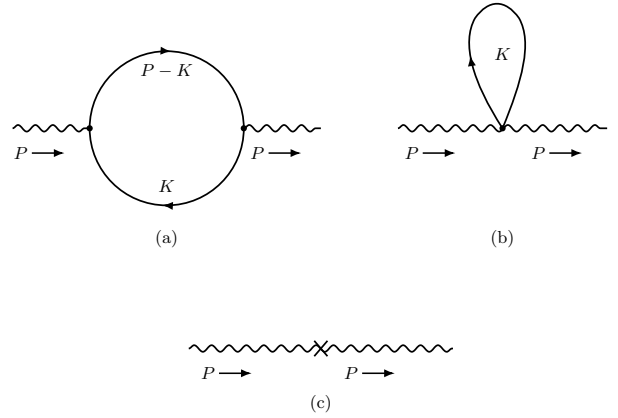
\begin{figure}[h]
\centering
\resizebox{\columnwidth}{!}{%
\begin{tikzpicture}

\begin{scope}[shift={(0,0)}]
\draw[decorate, decoration={snake,amplitude=1.7pt,segment length=9pt}, line width=0.9pt]
(0,0) -- (1.4,0);
\fill (1.4,0) circle (1.6pt);
\fill (4.2,0) circle (1.6pt);
\draw[decorate, decoration={snake,amplitude=1.7pt,segment length=9pt}, line width=0.9pt]
(4.2,0) -- (5.6,0);

\draw[
  line width=0.9pt,
  postaction={decorate},
  decoration={markings,mark=at position 0.52 with {\arrow{Latex[length=2mm]}}}
] (1.4,0) arc[start angle=180,end angle=0,radius=1.4];

\draw[
  line width=0.9pt,
  postaction={decorate},
  decoration={markings,mark=at position 0.52 with {\arrow{Latex[length=2mm]}}}
] (4.2,0) arc[start angle=0,end angle=-180,radius=1.4];

\draw[-{Latex[length=2mm]}, line width=0.8pt] (0.35,-0.45) -- (0.95,-0.45);
\draw[-{Latex[length=2mm]}, line width=0.8pt] (4.65,-0.45) -- (5.25,-0.45);

\node at (0.15,-0.45) {$P$};
\node at (4.45,-0.45) {$P$};
\node at (2.8,1.05) {$P-K$};
\node at (2.8,-1.05) {$K$};
\node at (2.8,-2.0) {(a)};
\end{scope}

\begin{scope}[shift={(7.0,0)}]
\draw[decorate, decoration={snake,amplitude=1.7pt,segment length=9pt}, line width=0.9pt]
(0,0) -- (3.8,0);
\fill (1.9,0) circle (1.6pt);

\draw[
  line width=0.9pt,
  postaction={decorate},
  decoration={markings,mark=at position 0.72 with {\arrowreversed{Latex[length=2mm]}}}
]
(1.9,0)
.. controls (2.25,0.85) and (2.45,1.55) .. (2.2,2.0)
.. controls (1.95,2.35) and (1.65,2.35) .. (1.4,2.0)
.. controls (1.15,1.55) and (1.35,0.85) .. (1.9,0);

\draw[-{Latex[length=2mm]}, line width=0.8pt] (0.45,-0.45) -- (1.05,-0.45);
\draw[-{Latex[length=2mm]}, line width=0.8pt] (2.75,-0.45) -- (3.35,-0.45);

\node at (0.25,-0.45) {$P$};
\node at (2.55,-0.45) {$P$};
\node at (1.90,1.35) {$K$};
\node at (1.9,-2.0) {(b)};
\end{scope}

\begin{scope}[shift={(3.2,-4.0)}]
\draw[decorate, decoration={snake,amplitude=1.7pt,segment length=9pt}, line width=0.9pt]
(0,0) -- (4.8,0);

\draw[line width=0.9pt] (2.28,0.14) -- (2.52,-0.14);
\draw[line width=0.9pt] (2.28,-0.14) -- (2.52,0.14);

\draw[-{Latex[length=2mm]}, line width=0.8pt] (0.55,-0.45) -- (1.15,-0.45);
\draw[-{Latex[length=2mm]}, line width=0.8pt] (3.15,-0.45) -- (3.75,-0.45);

\node at (0.35,-0.45) {$P$};
\node at (2.95,-0.45) {$P$};
\node at (2.4,-1.0) {(c)};
\end{scope}

\end{tikzpicture}%
}
\caption{
One-loop vacuum-polarization contributions.
(a) Contribution from $V_{1}$.
(b) Contribution from $V_{2}$.
(c) Measure-factor contribution.
}
\label{fig:vacpol_loops_manual}
\end{figure}

So in our framework the presence of the ultraviolet scale does break exact conformal invariance but it does not destroy gauge invariance or the infrared Maxwell limit.

There are some Conceptual difference between the two frameworks, for example in the Kempf--Mangano--Mann framework the new microscopic scale appears directly in the canonical algebra and as a result the short-distance structure of the theory is altered already at the level of phase-space kinematics so any field-theoretic extension then inherits a scale at the algebraic level and exact conformal invariance is lost for that reason.

Where in the present framework the canonical commutator is not deformed but instead we keep the Lorentz-covariant spacetime manifold and replace strictly local observables by non-local regulated observables of the form \eqref{eq:OFdef}. The same ultraviolet scale again breaks exact conformal invariance but the breaking is now implemented through a gauge-covariant function of the d'Alembertian~\cite{ThompsonCovariance}. This emphasis on preserving spacetime symmetry at the quantum level is not merely aesthetic as in gauge theory the fate of the Poincar\'e algebra can distinguish inequivalent quantization procedures so maintaining the symmetry structure is part of maintaining the physical content of the theory, as explained very well by Richard Epp and Gabor Kunstatter~\cite{EppKunstatter1994}. Historically this sits in the broader development of gauge invariance from Weyl onward~\cite{Weyl1918,Weyl1929,JacksonOkun2001}. So the gauge sector remains consistent, Ward identities survive, and ordinary Maxwell theory is recovered in the infrared~\cite{Ward1950,Takahashi1957,Schwinger1951,Moffat1990FiniteNonlocalGauge,EvensMoffatKleppeWoodard1991,CoteFaraoniGiusti2019}.

The conclusion is thus that our theory is not exactly conformally invariant but it does not suffer from the same structural problem as a noncovariant minimal-length deformation. What is lost is exact scale invariance and what is preserved is Lorentz covariance, gauge covariance, and the infrared Maxwell limit.

\section{On Locality and the Structure of Spacetime}

Now the next logical question one might note is what the derived uncertainty relation says about QFT and spacetime and we will comment on this briefly. The first thing that is immediately noticeable is that locality becomes quantitative as in strictly local QFT microcausality is binary~\cite{Haag1992}:
\begin{equation}
[O_1(f),O_2(g)]=0,
\end{equation}
for spacelike-separated supports. In the nonlocal theory strict locality is softened as regulated observables possess noncompact tails and their commutators at spacelike separation are suppressed rather than identically zero~\cite{BorchersPohlmeyer1968,ThompsonAMC,ThompsonSMatrix}, so we have found that locality becomes quantitative where exact point supported separation is replaced by exponentially accurate separation once the invariant distance exceeds a few times $L_M$.

With the first point comes the second, and we see that spacetime points cease to be observables, this follows from the modified uncertainty relation implies that no detector can localize a degree of freedom to a region smaller than order of the Moffat length $L_M$. This does not mean that the manifold description of spacetime disappears rather the correct statement is subtler and tells us that the mathematical spacetime manifold remains continuous, Lorentz covariance and translation invariance remain intact, pointlike localization is no longer a physically realizable observable notion below $L_M$ and are no longer fundamental~\cite{Wightman1962,Busch1999,Malament1996,ThompsonOperational}.

So to state is simply is that spacetime remains a continuum in the geometric sense, but becomes finite-resolution in the physical sense.

An interesting feature is that microcausality is emergent in the infrared, where the ordinary local limit is recovered when $E_M\to \infty$, or equivalently when probes have characteristic momentum scales much smaller than $E_M$. In that regime where:
\begin{equation}
\Delta X_{\mathrm{NL}} \sim \frac{1}{2\Delta P},
\end{equation}
the kernel collapses toward a delta distribution, and exact microcausality is recovered. So this shows us that strict local QFT is the infrared effective limit of the nonlocal ultraviolet theory.

Building off the previous points this shows there is no need for discrete spacetime or broken Lorentz symmetry as the present framework does not force a lattice, a preferred frame, or an explicit breaking of Lorentz invariance. The regulator is built from an entire function of the Lorentz-covariant d'Alembertian~\cite{Moffat1990FiniteNonlocalGauge,MoffatThompsonReg}. So the ultraviolet modification is not a discretization of spacetime but a Lorentz-covariant smearing of physically accessible observables.

So with all these points, what is the interpretation for our universe you may ask? Well at the broadest level I would say the relation derived here suggests that the universe is described in the ultraviolet not by a collection of physical spacetime points but instead by a nonlocal structure with an intrinsic scale. Observed geometry, observed particle localization, and observed causal separation are all emergent coarse-grained notions below that scale. In this sense the nonlocality scale $L_M$ plays a role analogous to a shortest meaningful length, while preserving the continuum symmetries observed at accessible energies.

\section{Conclusion}

We have presented the nonlocal uncertainty principle from the actual observable structure of entire function regulated nonlocal quantum field theory. The essential statement is the exact variance-addition law which when we combined it with the local Heisenberg inequality gives a new uncertency relation and under the infrared expansion reproduces the usual Heisenberg relation plus a small nonlocal correction term, while the ultraviolet limit yields a minimal localization length of order $E_M^{-1}$.

The result that we found tells us that the ultraviolet theory remains a Lorentz covariant continuum where pointlike localization is not a fundamental observable notion, but what is fundamental is the nonlocal algebra of regulated observables and the finite-resolution detector effects it induces. The usual point-like picture of spacetime and exact microcausality are then recovered as infrared limits but are not fundamental like in the overlocalized quantum field theory.

Broadly speaking this shift away from fundamentally pointlike observables is consistent with some quantum gravitational approaches in which the important degrees of freedom are organized by subregion boundaries and their associated symmetry generators rather
than solely by naive bulk local variables such as those found in~\cite{Dupuis:QGSymmetry3D,FewsterRejznerAQFT,BuchholzFredenhagenAQFT,
PapageorgiouFraserMeasurement,FewsterVerchMeasurement,FalconeContiLocalization}.

A natural next step is to seek an experimental test of the non-local uncertainty relation derived in this work. Rather than probing the theory first through high-energy scattering I believe that a faithful test would be a localization experiment in which a matter-wave packet is prepared with tunable momentum width and its spatial profile is measured with high precision. The key question would be whether the observed localization width continues to decrease according to the ordinary local Heisenberg expectation or instead saturates at a nonzero minimal scale set by the intrinsic non-locality length. In physical terms such an experiment would test the central claim of the present framework, that the ultraviolet modification is not a breakdown of Lorentz covariance or quantum mechanics but a change in what can be resolved.

Another possible experimental test, if we extend to make momentum nonlocal, would be squeezed states. The reason is that squeezed states are designed precisely to redistribute uncertainty between conjugate variables, so they can therefore provide us with a direct way to test whether the ordinary local phase-space ellipse can be squeezed indefinitely in one direction, or whether it saturates at an intrinsic nonlocal width and how the uncertainty relation may scale. We could also look at how the uncertainty scales with the squeezed states. But the exploration of these ideas will appear in a follow up paper that is currently underway.

\section{Acknowledgments}
I would like to thank  my supervisor John Moffat, my friends and colleagues Hilary Carteret and Arvin Kouroshnia, and Robert Mann for helpful discussions on non-local quantum field theory, locality, and measurement in relativistic quantum theory.

\end{document}